\documentclass[aps,prl,showpacs,twocolumn,superscriptaddress]{revtex4-1}
\usepackage{amssymb}
\usepackage{amsmath}
\usepackage{graphicx}
\usepackage{amsfonts}
\usepackage{hyperref}
\usepackage{color}
\usepackage{epsfig}
\usepackage{bbm}

\usepackage[normalem]{ulem}

\newcommand{\bra}[1]{\ensuremath{\left\langle #1\right|}}
\newcommand{\ket}[1]{\ensuremath{\left|#1\right\rangle}}

\newcommand{\expect}[3]{\ensuremath{\left\langle{#1}\right|\!{#2}\!\left|{#3}\right\rangle}}

\newcommand{\rme}{\ensuremath{\mathrm{e}}}
\newcommand{\rmi}{\ensuremath{\mathrm{i}}}

\newcommand{\bea}{\begin{eqnarray}}
\newcommand{\eea}{\end{eqnarray}}

\def\id{{\rm 1\kern-.22em l}}
\newcommand{\rhoaxi}{\rho^{\mathrm{axi}}}

\DeclareMathOperator{\Tr}{Tr}

\begin{document}
\title{
Quantifying entanglement of maximal dimension in bipartite 
mixed states
      }
\author{Gael Sent\'is}
\affiliation{Departamento de F\'{i}sica Te\'{o}rica e Historia de la Ciencia, Universidad del Pa\'{i}s Vasco UPV/EHU, E-48080 Bilbao, Spain}
\author{Christopher Eltschka} 
\affiliation{Institut f\"ur Theoretische Physik, Universit\"at Regensburg, D-93040 Regensburg, Germany}
\author{Otfried G\"uhne}
\affiliation{Naturwissenschaftlich-Technische Fakult\"at, Universit\"at Siegen, 
57068 Siegen, Germany}
\author{Marcus Huber} 
\affiliation{F\'{i}sica Te\`{o}rica, Informaci\'{o} i Fen\`{o}mens Qu\`{a}ntics, 
Universitat Autonoma de Barcelona, ES-08193 Bellaterra (Barcelona), Spain
            }
\affiliation{Group of Applied Physics,  University of Geneva,  1211 Geneva 4,  Switzerland}
\affiliation{Institute for Quantum Optics and Quantum Information (IQOQI), Austrian Academy of Sciences, 
A-1090 Vienna, Austria}
\author{Jens Siewert}        
\affiliation{Departamento de Qu\'{i}mica F\'{i}sica, Universidad del Pa\'{i}s Vasco UPV/EHU, E-48080 Bilbao, Spain}
\affiliation{IKERBASQUE Basque Foundation for Science, E-48013 Bilbao, Spain
            }
\begin{abstract}
The Schmidt coefficients capture all entanglement properties of
a pure bipartite state and therefore determine its usefulness for
quantum information processing. While the quantification of the corresponding
properties in mixed states is important both from a theoretical and
a practical point of view, it is considerably more difficult, and
methods beyond estimates for the concurrence are elusive. 
In particular this holds for a quantitative assessment of the most
valuable resource, 
the forms of entanglement that can only exist in high-dimensional systems.
We derive a framework for lower bounding the appropriate measure of 
entanglement, the so-called {\em G-concurrence}, 
through few local measurements. Moreover, we show that these bounds have
relevant applications also for multipartite states.
\end{abstract}
\maketitle
%
%
%
Understanding the nature and operational uses of entanglement 
constitutes one of the key challenges of quantum information theory. 
While most algorithms that allow for a provable advantage with respect to 
classical computation exhibit this ubiquitous feature of quantum systems, 
it is not entirely clear whether it actually is required for the 
promising field of quantum computation and simulation to outperform 
their respective classical counterparts. 

Consequently much effort has been invested in understanding the 
interplay between entanglement structure and resource properties of 
multipartite quantum 
states~\cite{Horodecki2009,GuehneToth2009,ES2014,MHetal2016}. 
One of the key results for 
computing with pure quantum states is the fact that, in order to go beyond 
the classical realm, a large dimension of entanglement is required 
while indeed any actual 
continuous measure of entanglement can be rather 
small~\cite{vdNest2013}.
Whether or how this statement translates to realistic conditions, 
{\em i.e.}, mixed-state quantum computing, is not at all clear. 
Here one could imagine a speed-up without any entanglement present at all, 
or, on the contrary, the need for high-dimensional entanglement in a more 
robust sense. However, it appears intuitively clear that mixed states 
with substantial overlap to states, whose resource content is 
exponentially hard to simulate classically, 
continue to be sufficient.

To answer such questions and to ultimately gain a deeper insight into 
the very nature of entanglement one would need a thorough quantification 
of all possible features of mixed-state entanglement. 
The sheer complexity of this task makes general solutions unlikely 
(recall that even deciding whether or not a given state is entangled is 
an NP-hard problem \cite{Gharibian2010}). 

A first interesting step in this direction could be the quantitative 
characterization of high-dimensional entanglement, {\em i.e.}, the
most expensive resource in bipartite systems.
One of the paradigmatic measures for the dimensionality of entanglement is
the Schmidt number of mixed quantum states \cite{Terhal2000}, for which various methods of certification exist~\cite{Sanpera-Schmidt,Brunner-Dimension,DadaNatPhys}.
However, the Schmidt number in itself is not entirely significant,
as even the highest possible dimensionality can lie in the vicinity of 
completely separable states~\cite{Goldbart2003}.
A robust quantification of mixed-state 
entanglement dimensionality can be made by using continuous measures of 
entanglement dimension 
which possibly bear 
also an operational meaning, 
beyond the question of mere computability. 
%
For bipartite entangled states the natural candidate for this purpose
is the family of concurrence monotones introduced by Gour~\cite{Gour2005}.
For a $d\times d$-dimensional
system, there are $d-1$ such monotones $k=2,\ldots,d$. 
The $k$th  concurrence monotone (which we will call for short
$k$-concurrence) vanishes
for a given state if its Schmidt number does not exceed $k-1$.
The usual concurrence~\cite{SMFei2001,Rungta2001}
coincides with the 2-concurrence in this family (up to a 
normalization constant). The measure for $k=d$
quantifies to which extent the maximum Schmidt number is contained in
a state and is usually termed $G$-concurrence.
While there exist various bounds for the usual concurrence~\cite{Fei2005,Fei2010,Ma2011,Rafsanjani2012,Ma2012,ETS2015,Mintert2005}, 
there are no mixed-state bounds for any of the other concurrences
monotones,
in particular not for the $G$-concurrence. Such a bound would go beyond giving an answer to 
the question  whether or not a state contains
entanglement of maximum dimensionality.

This is exactly what we achieve 
in this article: We first derive a general 
method how this measure can be efficiently lower bounded by using nonlinear 
witness techniques, allowing for a mixed-state quantification of 
$G$-concurrence in an experimentally feasible way. 
Furthermore, we find the exact solution for the $G$-concurrence of
the so-called axisymmetric states~\cite{ES2013}, 
a highly symmetric two-parameter family of $d\times d$ mixed states.
This solution provides the basis for a simple method to find lower bounds
to the $G$-concurrence of arbitrary mixed states.
As a byproduct, it also allows us to find lower bounds to the distance between the state of interest and the set of separable states, and, more generally, to any set of states with bounded Schmidt number.

{\em Nonlinear G-concurrence witness.}---
We commence by a brief definition of the relevant concepts, 
before going to our first main theorem.
For pure quantum states 
$\ket{\psi}\in \mathbb{C}_A^d\otimes\mathbb{C}_B^d$,
$\ket{\psi}=\sum_{jk}c_{jk}|jk\rangle$ the $G$-concurrence is 
defined as the $d$th root of the product of the $d$ eigenvalues 
of the marginal~\cite{Gour2005}. 
%
Denoting the Schmidt coefficients of the state as $\lambda_j\geqq 0$
  ({\em i.e.}, $\ket{\psi} = \sum_j \lambda_j \ket{a_j b_j}$; 
  consequently the
  marginal eigenvalues are $\lambda_j^2$), we can define
\begin{align}
C_G(|\psi\rangle):=d(\lambda_1\lambda_2\cdots\lambda_d)^{\frac{2}{d}}
\,,
\end{align}
so that $0\leqq C_G(|\psi\rangle)\leqq 1\,\forall\,|\psi\rangle$. 
The extension to mixed states is straightforward via the 
convex roof~\cite{Uhlmann1998}
\begin{align}
C_G(\rho):=\min_{\{p_k, \psi_k\}}\sum_jp_jC_G(|\psi_j\rangle)\, ,
\end{align}
where the minimum is taken over all pure-state decompositions
$\rho=\sum_k p_k \ket{\psi_k}\!\bra{\psi_k}$.

The idea here is to derive a tight lower bound for the 
$G$-concurrence of pure states in a form that admits a straightforward 
extension to a nonlinear witness lower bound for mixed states. 
In spirit this work follows Refs.~\cite{Guehne2010,Huber2010,Ma2011},
that is, if a state does {\em not} belong to a certain entanglement class,
the modulus of the offdiagonal elements cannot exceed a certain 
monotonically increasing function of the diagonal elements.
Using elementary algebra we arrive at our first main result (the proof is given in the Appendix),
\begin{align}
C_G(\rho)\geqq B_G(\rho)=\sum_{j, k}((d-2)\delta_{jk}+1)|\langle jj|\rho|kk\rangle|-\nonumber\\ d\sum_{\sigma\neq\id}(\text{Tr}(\bigotimes_{j=0}^d|j\sigma(j)\rangle\langle j\sigma(j)|\rho^{\otimes d})^{\frac{1}{d}})\ \ ,
\label{lowbound1}
\end{align}
where $\sum_{\sigma\neq\id}$ denotes the sum over all permutations of the 
levels of party $B$, excluding the identical permutation.
This general lower bound is both surprisingly simple and transparent. It is
expressed 
via density matrix elements and requires the knowledge of only $O(d^2)$ 
out of the $d^4-1$ elements. 
While Eq.~\eqref{lowbound1} is written in terms of $d\times d$-dimensional 
systems, it is obvious that
the bound can directly be applied also to the different bipartitions of 
multipartite systems, 
as we will see in the example below. 
%
%
Before we proceed with a more detailed discussion let us briefly comment 
that the bound (\ref{lowbound1}) is tight at least for all maximally 
entangled states of dimension $d$, {\em i.e.},
 $B_G(|\Phi_d\rangle)=1$, 
where $|\Phi_d\rangle:=\frac{1}{\sqrt{d}}\sum_{j=0}^{d-1}|jj\rangle$.\\

By investigating the noise resistance we find that 
the worst possible kind of noise for our 
bound is white noise, as it maximally affects the 
negative terms in the bound.
 For dimension $d=3$, {\em e.g.}, we can study the white-noise tolerance
by considering the state 
$\rho(p)=p|\Phi_d\rangle\langle\Phi_d|+ \frac{1-p}{d^2}\id_{d^2}$. 
Inserting this state into our bound we find that it can reveal the presence 
of $3$-concurrence down to 
$p=\frac{2}{3}$ (which is close to the exact
value $p=\frac{5}{8}$, see below). 
However, for higher dimensions the noise resistance of the $G$-concurrence 
decreases rapidly.
Possibly the quality of the bound~\eqref{lowbound1} can be improved by 
finding different estimates of the $G$-concurrence for pure states.
%

Interestingly, the nonlinear witness Eq.~\eqref{lowbound1} is not the
only possibility to estimate the quality of high-dimensional
entanglement in a mixed bipartite state. In the following, we first
describe the exact solution of $C_G(\rho)$ for certain
symmetric states. By means of this solution we can achieve
an independent lower bound for arbitrary $d\times d$ states.
%

{\em Exact solution for axisymmetric states.}---
Families of highly symmetric states often allow for an exact solution of
entanglement-related 
problems~\cite{Vollbrecht2000,Vollbrecht2001,Buchholz2016}. 
Here we consider the axisymmetric states,
a two-parameter family of $d\times d$-dimensional 
mixed states~\cite{ES2013,ETS2015}. 
%
They comprise all mixed states that have the same
symmetries as $\ket{\Phi_d}$, that is, (i) permutation symmetry of the 
qudits, (ii) invariance under simultaneous exchange of two levels for
both parties, that is, 
$\ket{j}_A\leftrightarrow\ket{k}_A$, $\ket{j}_B\leftrightarrow\ket{k}_B$, 
and (iii) symmetry under simultaneous  local  phase rotations
%
%
\begin{align*}
          \;\;\;\; \;\; \;\;\;\;V(\varphi_1,\varphi_2,\ldots,\varphi_{n-1})=
          \rme^{\rmi\sum \varphi_j \mathfrak{g}_j}\otimes
          \rme^{-\rmi\sum \varphi_j \mathfrak{g}_j}\ \ .
\end{align*}
Here, $\mathfrak{g}_j$ are the $(d-1)$ diagonal generators
of SU($d$).

%
The axisymmetric states can be written as mixtures of three states
\begin{align}\label{rhoaxi}
      \rho^{\mathrm{axi}}\ =\ & p \ket{\Phi_d}\!\bra{\Phi_d}+ (1-p) 
                  \left[q\tilde{\rho}_1+(1-q)\tilde{\rho}_2\right]\ \ ,
\nonumber\\
 \tilde{\rho}_1\ =\ &  \frac{1}{d-1}\sum_{m=1}^{d-1} |\Psi_d^{(m)}\rangle\langle\Psi_d^{(m)}|
\ \ ,
\nonumber\\
      \tilde{\rho}_2\ =\ & \frac{1}{d(d-1)}\sum_{j\neq k}
                                       \ket{jk}\!\bra{jk} 
\end{align}
where $0\leqq p,q\leqq 1$, and $|\Psi_d^{(m)}\rangle:= \tfrac{1}{\sqrt{d}} \sum_j \exp(i2\pi j\, m/d)\ket{jj}$ is a maximally entangled state with phase factors. They 
can be represented by a triangle. 
Remarkably it was found that Schmidt-number related entanglement
properties are affine functions of
the fidelity of $\rhoaxi$ with the maximally entangled state 
$F=\Tr\left(\rhoaxi\ket{\Phi_d}\!\bra{\Phi_d}\right)$.
For example, the borders of the Schmidt-number
classes are lines of constant fidelity $F$ (for $F\geqq\frac{1}{d}$).
In Ref.~\cite{ETS2015} it was shown that also the 2-concurrence of $\rhoaxi$
is an affine function of $F$, namely     
$C_2(\rhoaxi)=\sqrt{\frac{2d}{d-1}}\left(F-\frac{1}{d}\right)$ for $F\geqq 
\frac{1}{d}$.
By using the methods from Refs.~\cite{Vollbrecht2000} and 
\cite{Vollbrecht2001} 
we show that the exact $G$-concurrence for axisymmetric states is
\begin{align}
            C_G(\rhoaxi)\ =\ \max[ 1-d(1-F), 0  ] \ \ .   
\label{lowbound2}
\end{align}

In order to prove Eq.~\eqref{lowbound2}, one first notes
that for symmetric mixed states it suffices to minimize $C_G$ 
for pure states $\psi$ as a function of the Schmidt coefficients
under the constraint of fixed fidelity $F=\left|\langle \psi | \Phi_d\rangle
                          \right|^2=\Tr\left(\rhoaxi \ket{\psi}\!\bra{\psi}\right)$
and to convexify the resulting function (cf.\ Ref.~\cite{Vollbrecht2001})
\begin{align}
     C_G(\rhoaxi)\ =\ {\rm co} C_G(\psi)\ \ .
\label{VoWe}
\end{align}
(here, ${\rm co}C_G(\psi)$ denotes the convex hull). 
In complete analogy with the approach in Ref.~\cite{Vollbrecht2000}
one finds that the problem effectively depends only on a {\em single}
parameter, the fidelity $F$:
\begin{align}
      C_G(F)\ =\ d\left(\alpha\beta^{d-1}\right)^{\frac{2}{d}} \,,
\label{pure}
\end{align}
where
\begin{align*}
     \alpha\ =\ & \frac{1}{\sqrt{d}}\left(\sqrt{F}-\sqrt{d-1}\sqrt{1-F}
                                    \right)\ \ ,\ \ \
     F\geqq\frac{d-1}{d}\ ,
\\   \beta\ =\ & \frac{1}{\sqrt{d}}\left(\sqrt{F}+
                            \frac{\sqrt{1-F}}{\sqrt{d-1}}\right)\ \ .
\end{align*}
In the Appendix 
we present more details of this derivation.
Moreover, we prove that the function in Eq.~\eqref{pure} is concave
such that its convex hull is the affine function~\eqref{lowbound2}.
We show the result in Fig.~1 for $d=4$.
\begin{figure}[h]
  \centering
 \includegraphics[width=.97\linewidth]{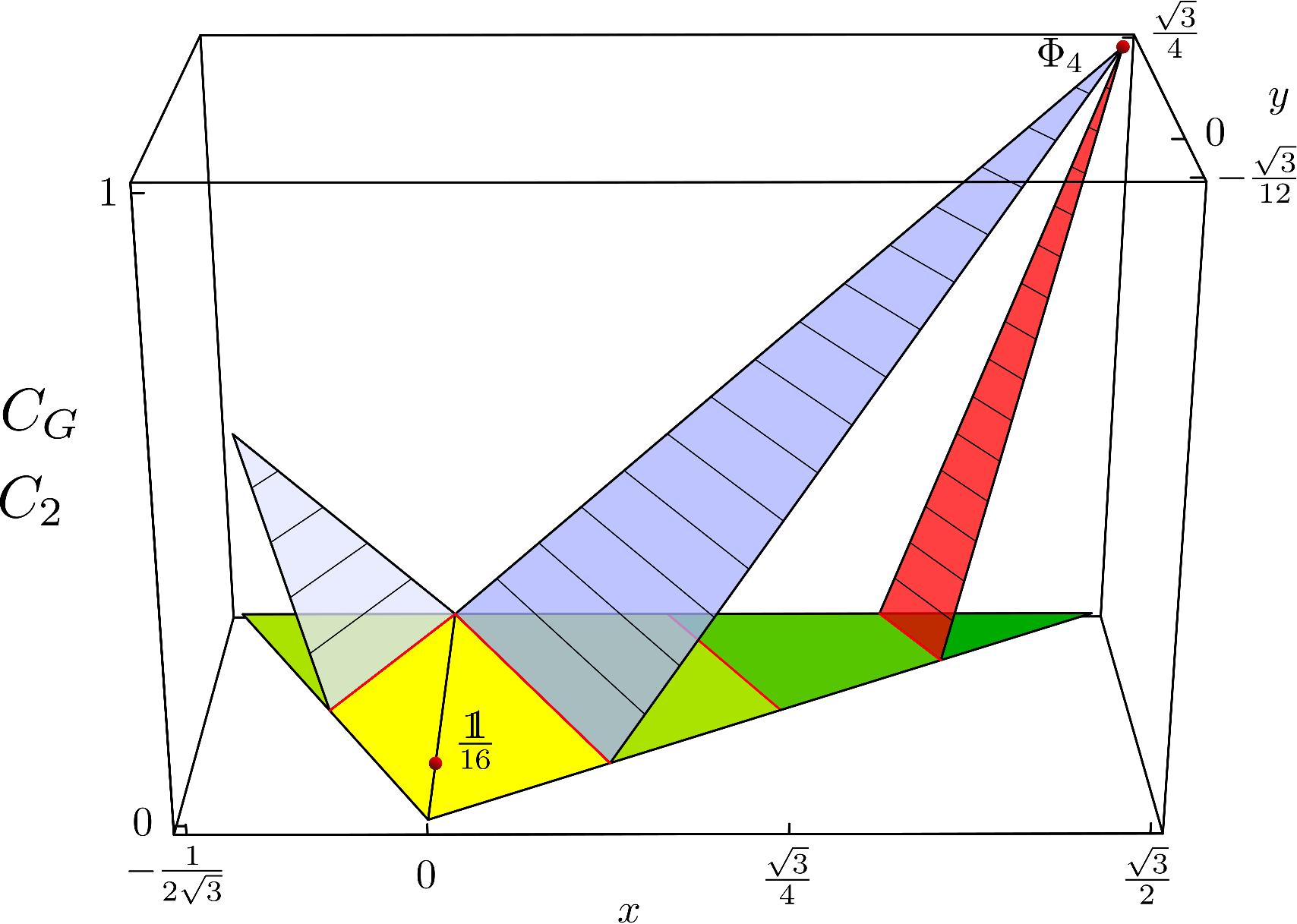}
  \caption{
The $G$-concurrence (red) and the 2-concurrence (light blue)
for $4\times 4$ axisymmetric states. The 2-concurrence (for pure states)
is defined here as $C_2=\sqrt{\frac{d}{d-1}\left(1-\Tr\rho_A^2\right)}$,
where $\rho_A$ denotes the reduced state of party $A$.
In the plane we show the axisymmetric states and
the borders between the Schmidt-number classes (red solid lines)
which, for $x>0$ are lines
of constant fidelity $F=\Tr\left(\rhoaxi \ket{\Phi_4}\!\bra{\Phi_4}\right)$. 
Here,
$x$ and $y$ are the appropriate coordinates to parametrize the 
axisymmetric states in a geometry that corresponds to the 
Hilbert-Schmidt metric~\cite{ES2013,ETS2015}.
          }
\end{figure}

{\em Arbitrary states.}---
The exact solution for axisymmetric states is interesting
not only from a mathematical point of view. We can use it
to obtain a lower bound on $C_G(\rho)$ for arbitrary states $\rho$
by noting that the average over all the operations in the
group of axisymmetries $\mathcal{V}$ (``twirling'') applied to 
$\rho$ represents a projection
\begin{align*}
   \mathbb{P}^{\mathrm{axi}}(\rho)\ =\ \int\!\! \mathrm{d} 
                                        \mathcal{V}\ \mathcal{V}\rho
                                          \mathcal{V}^{\dagger}
\end{align*}
into the axisymmetric states~\cite{Vollbrecht2001}.
On the other hand, averaging over the operations $\mathcal{V}$
can only {\em reduce} the entanglement, so that for 
  $\rho^{\mathrm{axi}}(\rho):=\mathbb{P}^{\mathrm{axi}}(\rho)$ we have
a lower bound~\cite{Vidal2000,Horodecki2009}
\begin{align}
            C_G[\rhoaxi(\rho)]\ \leqq\ C_G(\rho)\ \ .
\label{symm}
\end{align}
This bound, which explicit form is given in Eq.~\eqref{lowbound2}, has been recently proven to hold also for fidelity parameters taken with respect to arbitrary states~\cite{Zhang2015}.
The components of the symmetrized state $\rhoaxi(\rho)$ are easily
obtained via the relations
\begin{subequations}
\begin{align}
   \rho^{\mathrm{axi}}_{jk,jk} &  = \frac{1}{d}\left[
                                      \delta_{jk} \sum_a \rho_{aa,aa} 
                                  +
                                      \frac{1-\delta_{jk}}{d-1}
                      \sum_{a \neq b} \rho_{ab,ab}
                                                \right]
\\
   \rho^{\mathrm{axi}}_{jk,lm} &  = 
                   \frac{\delta_{jk}\delta_{lm}(1-\delta_{jl})}{d(d-1)} 
                     \sum_{a>b} \left(\rho_{aa,bb}+\rho_{bb,aa}\right)  
\  .
\label{eq:symmmatels}
\end{align}
\end{subequations}
The symmetrization requires some care since one may lose all the 
entanglement by inappropriately choosing the local bases. By exploiting
local unitary invariance of the $G$-concurrence,
$C_G(\left[U_A\otimes U_B\right]\rho
                   \left[U_A\otimes U_B\right]^{\dagger})=C_G(\rho)$
we may improve the bound by finding the best local unitaries before 
doing the projection~\eqref{symm} so as to achieve the largest
$C_G[\rhoaxi(\rho)]$. Clearly, this holds as well for the bound
in Eq.~\eqref{lowbound1}.

Indeed, both the bounds~\eqref{lowbound1} and~\eqref{symm} 
can be improved even further by observing that
the $G$-concurrence is an SL($d,\mathbb{C})^{\otimes 2}$ 
invariant~\cite{ES2014}. According to 
Verstraete {\em et al.}~\cite{Verstraete2003} an entanglement monotone
based on a local SL invariant is maximized on the so-called normal
form of the state. The hallmark of the normal form is that it has
maximally mixed local density matrices~\cite{Verstraete2003,Myrheim2006}. 
It can be found via an algorithm
described in~\cite{Verstraete2003}. Thus, an exact solution (or lower bound)
for a local SL invariant like $C_G$
over a family of symmetric states can be used to calculate a lower bound
of $C_G(\rho)$ for arbitrary states 
$\rho$ by the following procedure~\cite{ES2012-ScR}:
\begin{itemize}
\item find the normal form 
      $\rho^{\mathrm{NF}}(\rho)$ (in general not normalized to 1;
     re-normalization is not necessary because of the 
     homogeneity of $C_G$ of degree 1 in the density matrix); 
if the normal form vanishes the procedure terminates
and $C_G(\rho)=0$; 
\item apply optimal local unitaries to $\rho^{\mathrm{NF}}(\rho)$
      (as described above) which leads to $\tilde{\rho}^{\mathrm{NF}}$
      and do the projection 
      $\mathbb{P}^{\mathrm{axi}}\left(\tilde{\rho}^{\mathrm{NF}}\right)
      =: \tilde{\rho}^{\mathrm{NF}}_{\mathrm{axi}}$\ ;
\item read off the bound for the $G$-concurrence from this state
\begin{align}
      C_G(\rho)\ \geqq\ 
                        C_G\left(\tilde{\rho}^{\mathrm{NF}}_{\mathrm{axi}}
                           \right)\ \ .
\label{SLbound}
\end{align}
\end{itemize}
Clearly, in order to produce the normal form knowledge
of all the matrix elements of $\rho$ is required. Hence 
improving the bounds via  SL($d,\mathbb{C})^{\otimes 2}$
(as well as via unitary) optimization is more
expensive with respect to the experimental effort.

To conclude this section, we show how our results can also be used in some cases to guarantee that an arbitrary state $\rho$ has a finite distance to any set of states with bounded Schmidt number---in particular to the set of separable states---, thus rendering the computed bounds of $C_G(\rho)$ more meaningful and robust measures of the entanglement of $\rho$. Let $\mathcal S_k$ be the set of all states with Schmidt number $k<d$, $\mathcal S_k^{\rm axi}$ the set of Schmidt number $k$ states in the axisymmetric family, and $\rho^{\rm axi}$ the symmetrization of $\rho$. Then, the following inequality holds~\cite{footnote}:
\begin{equation}\label{ineq_distance}
\min_{\sigma\in{{\mathcal S}^{\rm axi}_k}} ||\rho^{\rm axi}-\sigma ||_{\rm HS} \le \min_{\sigma\in \mathcal{S}_k} || \rho - \sigma||_{\rm HS} \,.
\end{equation}
This inequality tells us that, given $\rho$, whenever its projection $\rho^{\rm axi}$ lies at a finite distance with respect to the closest axisymmetric state with Schmidt number $k$, we know that the distance between $\rho$ and the closest Schmidt number $k$ state is at least as large. A consequence of this is that any nonzero value of the bound $C_G[\rho^{\rm axi}(\rho)]$ rules out the possibility of $\rho$ being arbitrarily close to a separable state.

{\em Application of the bounds to multipartite states.}---
While the usefulness of our bounds for the characterization
of bipartite states is apparent, we would like to point out that
this is true also in the context of multi-party states. To this end,
let us consider a four-qubit cluster state
\begin{align*}
      \ket{\mathrm{Cl}_{ABCD}} = \frac{1}{2}\left(
                    \ket{0000}+\ket{0111}+\ket{1011}+\ket{1100}\right)\ .
\end{align*}
Each of the bipartitions $(AB)(CD)$, $(AC)(BD)$, and $(AD)(BC)$ may
be regarded as  a $4\times 4$ system where the state of $(AB)(CD)$
is of Schmidt rank 2 and the others have Schmidt rank 4.
Indeed, for the latter bipartitions the state is locally equivalent
to $\ket{\Phi_4}$ and has maximal $G$-concurrence, {\em e.g.},
$C_G(\ket{\mathrm{Cl}_{(AC)(BD)}})=1$.

Now we may ask how this resource behaves when noise is added to
$\ket{\mathrm{Cl}_{ABCD}}$.  We use the white-noise tolerance of the 
$G$-concurrence on a rank-4 bipartition as a model to answer this question.
Physically, this means we ask up to which 
admixture $w_4$ of white noise
any decomposition of the resulting state
$\rho_{ABCD}=(1-w) \ket{\mathrm{Cl}_{ABCD}}\!\bra{\mathrm{Cl}_{ABCD}}
+\frac{w}{16}\id_{16}$ contains a state of Schmidt
rank 4, {\em e.g.}, on the bipartition $(AC)(BD)$. The corresponding
fidelity is $F_4=\frac{3}{4}$ so that $w_4=\frac{4}{15}$.
This is a remarkable result, as it has to be contrasted with the
noise tolerance of genuine multipartite 
entanglement (GME) for this state, $w_4^{\mathrm{GME}}=\frac{8}{13}$
(cf.~Ref.~\cite{Guehne2011}). 
It shows, as expected, that a well-specified resource of multipartite entanglement behaves differently from GME.

This discussion is straightforwardly extended to linear cluster
states of large (even) number $N$ of qubits. The existence
of bipartitions with full Schmidt rank is one of their 
important properties~\cite{Hein2006}.
In that case, the maximum
Schmidt rank across the bipartitions is $d=2^{(N/2)}$, and hence
$w_N\simeq 2^{-(N/2)}$. Recall that for linear $N$-qubit cluster
states $w^{\mathrm{GME}}_N \simeq 1-(N/3)2^{-(N/3)}$~\cite{Jungnitsch2011}.
That is, while the noise tolerance of GME in large linear cluster states
is practically perfect, the maximum Schmidt-rank resource becomes 
exponentially fragile with increasing $N$.

{\em Conclusions.}---
We have presented 
two independent quantification methods for high-dimensional
entanglement, {\em i.e.}, of the resource characterized by
the maximum number of non-vanishing Schmidt coefficients,
in bipartite mixed states. This is achieved by estimates
of the $G$-concurrence via a nonlinear witness on the one hand,
and by an exact solution for axisymmetric states on the other hand.
Our nonlinear witness Eq.~\eqref{lowbound1} extends the possibility
to detect entanglement of Schmidt number 2~\cite{Huber2010}
to maximum Schmidt number $d$. At the same time,
this nonlinear witness~\eqref{lowbound1}, as well as the projection 
witness~\eqref{symm}, is quantitative~\cite{Eisert2007} 
and can be 
experimentally determined by measuring a number of observables
of order $d^2$ which is considerably smaller than $d^4-1$, the number of
all parameters of the state. This shows that the developed methods are suitable
also to provide a quantitative analysis of recent efforts at producing high-dimensionally entangled states in the lab. 
The fact that Schmidt numbers equal to the system dimension were certified e.g. in Refs.~\cite{exp1,exp2,exp3,exp4} implies that the respective $G$-concurrence will be nonzero and the data taken should suffice to apply our methods.
Due to the possibility of SL($d,\mathbb{C})^{\otimes 2}$ optimization,
entanglement detection through our approach is superior compared to
merely using an optimal Schmidt number witness. However, exploiting this possibility
requires complete knowledge of the state parameters.
Moreover, we have outlined how our methods can be applied also
in the investigation of multipartite entanglement. We have shown that
the resource of maximum Schmidt number across the bipartitions of
$N$-qubit cluster states is exponentially fragile with respect to
the admixture of white noise. 
In addition, we mention that, in principle,
it is 
possible to define a genuine multipartite $G$-concurrence
in analogy with Ref.~\cite{Ma2011} in order to 
quantitatively describe the Schmidt-number vectors of a
multipartite system~\cite{Huber2013-1,Huber2013-2}.
Finally, we note that similar techniques to the ones we develop in the first part could potentially be used to lower bound any quantity that can be expressed as a polynomial of state coefficients, such as other ${\rm SL}$ invariants~\cite{ES2014,Coffman2000}.

\emph{Acknowledgments.}---
We thank C. Jebarathinam for pointing out an error in Eq.~\eqref{rhoaxi} in the published version of this manuscript. This work was funded by 
ERC Starting Grant 258647/GEDENTQOPT (G.S.),
the German Research Foundation within 
SPP 1386 (C.E.), the FQXi Fund (Silicon Valley Community Foundation),
the German Research Foundation (DFG), and ERC Consolidator Grant 683107/TempoQ
(O.G.), by 
Austrian Science Fund (FWF) through the START project Y879-N27,
Swiss  National  Science  Foundation (AMBIZIONE Z00P2-161351),
Spanish MINECO (Project  No.\  FIS2013-40627-P),
and the   Generalitat   de   Catalunya   CIRIT, Project
No.  2014-SGR-966 (M.H.),
by Basque Government grant IT-472-10,
MINECO grants FIS2012-36673-C03-01,
              FIS2012-36673-C03-03  and  FIS2015-67161-P, 
and UPV/EHU program UFI 11/55 (G.S.\ and J.S.). 
The authors would like to thank 
G.\ T\'oth for stimulating discussions, and J.\ Fabian and K.\ 
Richter for their support.



\appendix

\section{APPENDIX}

\section{The nonlinear witness for the G-concurrence}

Given an arbitrary pure state $\ket{\psi}=\sum_{i,j} c_{ij} \ket{ij}$, its $G$-concurrence can be computed as
\begin{equation}
C_G(\ket{\psi}) = d \left(\lambda_1 \lambda_2 \cdots\lambda_d \right)^{2\over d} = d	|\det c|^{2\over d} \,,
\end{equation}
where the $\lambda_i$ are its Schmidt coefficients (i.e. \mbox{$\ket{\psi}=\sum_i \lambda_i \ket{\alpha_i \beta_i}$}), and $c$ is a $d\times d$ matrix with elements $c_{ij}$.
We can use the triangle inequality twice to lower bound the determinant as
\begin{eqnarray}
|\det c| &=& |\sum_\sigma sgn(\sigma) \prod_{i=1}^d c_{i \sigma(i)}| \\
&\geq& |\prod_{i=1}^d c_{ii}| - |\sum_{\sigma\neq \id} sgn(\sigma)\prod_{i=1}^d c_{i\sigma(i)}|\\
&\geq & |\prod_{i=1}^d c_{ii}| - \sum_{\sigma\neq \id}|\prod_{i=1}^d c_{i\sigma(i)}| \,,
\end{eqnarray}
where $\sigma=\id$ is the identity permutation. Now, let us rename $|\prod_{i=1}^d c_{ii}|\equiv X$ and $\sum_{\sigma\neq e}|\prod_{i=1}^d c_{i\sigma(i)}|\equiv Y$. 
For any two positive numbers $X,Y$, it is immediate to check that 
$$
f \equiv X^{2\over d}-Y^{2\over d} \leq |X-Y|^{2\over d} \equiv g \,.
$$
Indeed, if $X<Y$, the inequality is trivially satisfied. On the other hand, if $X \geq Y \geq 0$, we just have to look at the convexity of $f$ and $g$. At the extreme points of this interval, i.e. when $Y=0,X$, the inequality is saturated. To see what happens in between, we compute the second derivatives of $f$ and $g$ with respect to $Y$, for an arbitrary $X$:
\begin{eqnarray}
f''\equiv \frac{d^2 f}{dY^2} &=& \frac{2}{d} \left(1-{2\over d}\right)Y^{{2\over d}-2} \,,\\
g''\equiv \frac{d^2 g}{dY^2} &=& -{2\over d} \left(1-{2\over d}\right) \left(X-Y\right)^{{2\over d} -2} \,.
\end{eqnarray}
We readily see that $f''\geq 0$ and $g''\leq 0$ for $d\geq 2$, which means that $f$ is convex and $g$ is concave, thus $f\leq g$ and we can write 
%
\begin{equation}\label{ineqDet}
|\det c|^{2\over d}\geq |\prod_{i=1}^d c_{ii}|^{2\over d} - \sum_{\sigma\neq \id}|\prod_{i=1}^d c_{i\sigma(i)}|^{2\over d} \,.
\end{equation}

It will prove useful to further lower the bound by replacing the positive term in the r.h.s. of Eq.~\eqref{ineqDet} by a bilinear function of the coefficients $c_{ii}$, namely
\begin{equation}\label{ineqplus}
|\prod_{i=1}^d c_{ii}|^{2\over d} \geq \alpha \sum_{i\neq j} c_{ii} c_{jj}^* - \beta \sum_{i=1}^d |c_{ii}|^2 
\end{equation}
for some real coefficients $\alpha$ and $\beta$. In order to prove this new inequality, we begin by rewriting it as 
\begin{equation}
|\prod_{i=1}^d c_{ii}|^{2\over d} \geq \alpha |\sum_{i=1}^d c_{ii}|^2 -(\alpha+\beta)\sum_{i=1}^d |c_{ii}|^2 \,.
\end{equation}
Note that the modulus makes it completely independent on complex phases, hence we can consider the coefficients $c_{ii}$ to be real for the rest of this proof. Furthermore, the inequality is scale invariant, so we deliberatively fix $\sum_{i=1}^d c_{ii}^2=1$ and, once again, we rewrite
\begin{align}\label{ineqPmin}
P(\{c_{ii}\})^{2\over d} & -\alpha S^2+\alpha+\beta \nonumber\\
&\geq P_{\rm min}(S)^{2\over d}-\alpha S^2+\alpha+\beta \geq 0\,,
\end{align}
where we have defined $P(\{x_i\})\equiv\prod_{i=1}^d x_i$, and 
$P_{\rm min}(S) = \min_{\{x_i\}:\sum_i x_i^2 = 1 ; \sum_i x_i = S}P(\{x_i\})
$ for a set of $d$ arbitrary parameters $\{x_i\}\in[0,1]^d$.
It is clear that \mbox{$P_{\rm min}(S\leq \sqrt{d-1})=0$}, since one can choose \mbox{$x_d=0$} and still find a set $\{x_i\}_{i=1}^{d-1}$ that fulfils the required conditions. As we increase the value of $S$ above this threshold, the minimum should still be attained when $x_d$ is as close to zero as possible. Such minimal value of $x_d$ is directly obtained by solving the simpler minimization $\min_{\{x_i\}} x_d$, subject to the original constraints. This is a straightforward calculation via Lagrange multipliers. The corresponding Lagrangian is
$$
\mathcal{L} = x_d -\lambda \left(\sum_{i=1}^d x_i-S\right) + {\mu\over 2}\left(\sum_{i=1}^d x_i^2 -1\right) \,,
$$
where $\lambda,\mu$ are Lagrange multipliers, and its symmetry already tells us that all the $x_{i<d}$ have to be equally valued. We find $x_i=\lambda/\mu$ for $i<d$, and $x_d = (\lambda-1)/\mu$. The multipliers are 
$$
\lambda = \frac{\mu S+1}{d} \,,\quad \mu = \pm\sqrt{\frac{d-1}{d-S^2}} \,.
$$
We see from the form of $\mu$ that $S\leq \sqrt{d}$. In terms of $S$ and $d$, an extreme value of $\prod_i x_i$ is attained when the coefficients are
\begin{eqnarray}
x_{i<d} &=& \frac{S\pm \sqrt{(d-S^2)/(d-1)}}{d} \,,\\
x_d &=& \frac{S\mp \sqrt{(d-S^2)(d-1)}}{d} \,.
\end{eqnarray}
The first solution corresponds to the minimum.

We now prove Eq.~\eqref{ineqPmin} by solving for $\alpha$ and $\beta$ the more restrictive set of inequalities
\begin{eqnarray}
-\alpha S^2+\alpha+\beta \geq 0 \,,& \forall S<\sqrt{d-1}\label{ineqL1}\\
\!\!\!P_L(S)-\alpha S^2+\alpha+\beta \geq 0 \,,& \forall \sqrt{d-1}\leq S\leq \sqrt{d}\,,\label{ineqL2}
\end{eqnarray}
where $P_L(S)$ is a linear lower bound of $P_{\rm min}(S)^{2\over d}$ that is tight at the extreme values of the interval $\sqrt{d-1}\leq S\leq \sqrt{d}$. The function $P_L(S)$ exists if $P_{\rm min}(S)^{2\over d}$ is a fully concave function. One can readily check that this is indeed the case, for the equation
$$
P_{\rm min}^{''}(S)\equiv\frac{d^2 (P_{\rm min}(S)^{2\over d})}{d S^2} = 0
$$
has no solution in the relevant domain, hence there are no inflection points. Then one only has to observe the sign of an intermediate point, e.g. $P_{\rm min}^{''}(\sqrt{d-1/2})$. Such function approaches zero exclusively in the asymptotic limit $d\to\infty$, and it is negative for any other (finite) value of $d\geq 3$, hence $P_{\rm min}^{''}(\sqrt{d-1/2})$ is always negative and therefore $P_{\rm min}(S)^{2\over d}$ is a concave function of $S$, for any $d$. Taking into account that $P_{\rm min}(S\leq\sqrt{d-1})=0$ and $P_{\rm min}(\sqrt{d})=1/d$, we may then write $P_L(S)$ as
\begin{equation}
P_L(S)=\frac{S-\sqrt{d-1}}{d(\sqrt{d}-\sqrt{d-1})} \,.
\end{equation}

Finding values of $\alpha$ and $\beta$ such that Eqs.~\eqref{ineqL1} and \eqref{ineqL2} hold is straightforward.
By, e.g., demanding that Eq.\eqref{ineqL2} be tight for $S=\sqrt{d}$, we get rid of one parameter. We obtain $\beta=\alpha(d-1)-1/d$. Then, imposing that Eq.~\eqref{ineqL1} be tight for $S=\sqrt{d-1}$ yields $\alpha = 1/d$, and thus $\beta=1-2/d$. With these values of $\alpha$ and $\beta$, we can guarantee that Eq.~\eqref{ineqPmin} is satisfied for all $S$. 

Summing up, we use Eqs.~\eqref{ineqDet} and \eqref{ineqplus} to lower-bound the \mbox{$G$-concurrence} of an arbitrary bipartite pure state $\ket{\psi}$ as $C_G (\ket{\psi}) \geq C_G^\downarrow (\ket{\psi})$, where
\begin{equation}
C_G^\downarrow (\ket{\psi}) =   \sum_{i\neq j} c_{ii} c_{jj}^* - (d-2) \sum_{i=1}^d |c_{ii}|^2 - d\sum_{\sigma\neq \id}|\prod_{i=1}^d c_{i\sigma(i)}|^{2\over d} \,.
\end{equation}

The lower bound for the convex roof extension to mixed states, $C_G(\rho)$, hence follows:
\begin{widetext}
\begin{eqnarray}
C_G(\rho) &=& \min_{\{p_k,\psi_k\}}\sum_k p_k C_G(\ket{\psi_k}) \geq \min_{\{p_k,\psi_k\}} \sum_k p_k C_G^\downarrow (\ket{\psi_k}) \nonumber\\
&\geq& \min_{\{p_k,\psi_k\}} \sum_k p_k \sum_{i\neq j} c^k_{ii} c^{k\,*}_{jj} - \max_{\{p_k,\psi_k\}} \sum_k p_k\left((d-2) \sum_{i=1}^d |c^k_{ii}|^2 + d\sum_{\sigma\neq \id}|\prod_{i=1}^d c^k_{i\sigma(i)}|^{2\over d}\right) \nonumber\\
&=& \sum_{i\neq j} \expect{ii}{\rho}{jj} - (d-2) \sum_{i=1}^d \expect{ii}{\rho}{jj} - d\sum_{\sigma\neq \id}\max_{\{p_k,\psi_k\}}\sum_k p_k |\prod_{i=1}^d c^k_{i\sigma(i)}|^{2\over d} \nonumber\\
&\geq& \sum_{i\neq j} \expect{ii}{\rho}{jj} - (d-2) \sum_{i=1}^d \expect{ii}{\rho}{jj} - d\sum_{\sigma\neq \id} \left(\prod_{i=1}^d\expect{i\sigma(i)}{\rho}{i\sigma(i)}\right)^{1\over d} \,,
\end{eqnarray}
\end{widetext}
where for the last inequality we have used the subadditivity of the root function to write

\begin{align}
\sum_k p_k |\prod_{i=1}^d c^k_{i\sigma(i)}|^{2\over d} &\leq& \left(\sum_k p_k \prod_{i=1}^d |c^k_{i\sigma(i)}|^2\right)^{1\over d} \nonumber\\
&\leq& \left( \prod_{i=1}^d \sum_k p_k |c^k_{i\sigma(i)}|^2 \right)^{1\over d}\,.
\end{align}

\section{G-concurrence of axisymmetric states}

\subsubsection{Derivation of Eq.~\eqref{pure}}


As mentioned in the main text, the proof proceeds through a minimization of the entanglement
measure under consideration (here the $G$-concurrence) on pure states first.
Consider therefore $\ket{\psi}\in \mathbb{C}^d\otimes\mathbb{C}^d$
with its Schmidt decomposition $\ket{\psi}=\sum_{j=0}^d \lambda_j\ket{a_jb_j}$.
The minimization of $C_G(\ket{\psi})$ is under the condition
that the fidelity of $\ket{\psi}$ with the maximally entangled state
$\ket{\Phi_d}=\frac{1}{\sqrt{d}}\sum_{j=0}^d\ket{jj}$ be fixed,
$\left|\langle{\Phi_d}|\psi\rangle\right|^2=F_{\psi}$. 

We use a fact noted by Terhal and Vollbrecht~\cite{Vollbrecht2000}
that the largest value of the fidelity 
$F_{\psi}=\left|\langle{\Phi_d}|\psi\rangle\right|^2$ 
is obtained if
the Schmidt bases $\{a_j\}$, $\{b_k\}$ coincide with the computational basis.
With this choice of bases, the only remaining parameters are the
Schmidt coefficients $\lambda_j$, and
\begin{align}
      \sum_j \lambda_j\ =\ \sqrt{d F_{\psi}}\ \ .
\label{cond1}
\end{align}
We are interested in non-vanishing 
$C_G(\ket{\psi})$, therefore we can assume $\lambda_j\neq 0$.
Further, we have the normalization 
\begin{align}
     \sum_j \lambda_j^2\ =\ 1\ \ .
\label{cond2}
\end{align}
Since $x^{1/d}$ is monotonous, we can just minimize the product 
$\lambda_1^2 \lambda_2^2 \cdots \lambda_d^2$.
By introducing Lagrange multipliers $A$ and $B$ for the conditions above, 
we arrive at the equations
\begin{align}
     \lambda_j \left(\prod_{n\neq j} \lambda_n^2\ -\ B\right)
      \ = \ A\ \ \ ,\ \  j=1,\ldots,d\ \ .
\label{deqs}
\end{align}
We can use any two of the Eqs.~\eqref{deqs} to obtain
\begin{align}
   (\lambda_j-\lambda_k)\left(\lambda_j\lambda_k \prod_{n\neq j,k}
                        \lambda_n^2 \ -\ B\right)\ =\ 0\  \ ,
\label{d2eqs}
\end{align}
which can be satisfied if $\lambda_j=\lambda_k$ or 
$\lambda_j\lambda_k\prod_{n\neq j,k}\lambda_n^2=B$.

Now consider three coefficients $\lambda_j$, 
$\lambda_k$ and $\lambda_l$ such that
$\lambda_j\neq \lambda_k$ and
$\lambda_j\neq \lambda_l$.
From Eq.~\eqref{d2eqs} we have
$\lambda_j\lambda_k\lambda_l^2 \prod_{n\neq j,k,l}\lambda_n^2=B$
and
$\lambda_j\lambda_l\lambda_k^2 \prod_{n\neq j,k,l}\lambda_n^2=B$,
so that the left-hand sides must agree, from which it follows
that $\lambda_k=\lambda_l$. Therefore, there can be at most
two different values for the $\lambda_n$,
\begin{align}
   \lambda_1=\ldots =\lambda_m=\alpha\ \ ,\ \ \
   \lambda_{m+1}=\ldots =\lambda_d= \beta\ \ .
\end{align}
By inserting this into the conditions~\eqref{cond1}, \eqref{cond2}
one obtains
\begin{align}
    m\alpha+(d-m)\beta \ =\ & \sqrt{d F_{\psi}}
\\
    m\alpha^2+(d-m)\beta^2 \ =\ & 1\ \ .
\end{align}
For given $m$, those equations are solved by
\begin{align}
   \alpha_{\pm}\ =\  &\sqrt{\frac{F_{\psi}}{d}}\pm
                \sqrt{\frac{d-m}{m}}\sqrt{\frac{1-F_{\psi}}{d}}
\\
   \beta_{\pm}\ =\  &\sqrt{\frac{F_{\psi}}{d}}\mp
                \sqrt{\frac{m}{d-m}}\sqrt{\frac{1-F_{\psi}}{d}}\ \ .
\end{align}
For $m=0$ and $m=d$ there is no general solution; further, we see
that it is sufficient to consider $m<\frac{d}{2}$.
We know from the axisymmetric states that $F_{\psi}\geqq \frac{d-1}{d}$;
what remains to do is to determine the $m$ and the sign 
for the best lower bound. To this end, we check the derivatives
$\frac{{\rm d}\alpha}{{\rm d}m}$, $\frac{{\rm d}\beta}{{\rm d}m}$ 
and find that for the `+' sign both $\alpha$ and $\beta$ have
their minimum for maximum $m$, {\em i.e.}, $m=d-1$ (whereas the `-'
sign gives $m=1$). Both solutions can be mapped to one another.
By choosing the `-' sign and  $m=1$ we find Eq.~\eqref{pure}, as well as the corresponding $\alpha$ and $\beta$.

\subsubsection{Concavity of the pure-state minimum}
%
In this section we prove the concavity of Eq.~\eqref{pure}. We have 
\begin{align}        
      C_G\ & =\  d\left(\alpha\beta^{d-1}\right)^{\frac{2}{d}}
\\
           & =\  d\left(\frac{1}{B}-1\right)^{\frac{1}{d}} B 
                   (d-1)^{\frac{1}{d}-1} \,,
\end{align}
where
\begin{align}
               B\ :=\ & (d-1)\beta^2
\nonumber\\
         =\ & \frac{1}{d}\left(1+(d-2)F+2\sqrt{d-1}\sqrt{F(1-F)}\right)\ \ .
\label{B(F)}
\end{align}
The last line here is obtained
by substituting  $\alpha(F)$, $\beta(F)$, 
see Eq.~\eqref{pure}. 

The second derivative of $C_G$ with respect to $F$ is (up to constant
positive prefactors for fixed dimension $d>1$)
\begin{widetext}
\begin{align}
\frac{{\rm d}^2C_G}{{\rm d}F^2}\ \propto  &\
                             \frac{{\rm d}^2B}{{\rm d}F^2} 
                 \left(\frac{1}{B}-1\right)^{\frac{1}{d}-1}
                 \left(\frac{d-1}{d}\frac{1}{B}-1\right) +
\nonumber\\
       & + \left(\frac{{\rm d}B}{{\rm d}F}\right)^2
           \left(\frac{d-1}{d}\frac{1}{B}-1\right) \left(\frac{1}{d}-1\right)
           \left(\frac{1}{B}-1\right)^{\frac{1}{d}-2}\left(-\frac{1}{B^2}\right)
         +
\nonumber\\
       & + \left(\frac{{\rm d}B}{{\rm d}F}\right)^2 
                 \left(\frac{1}{B}-1\right)^{\frac{1}{d}-1}
                 \frac{d-1}{d}\left(-\frac{1}{B^2}\right)
\nonumber\\
        = &\
           \frac{1}{d^2 B^3}
           \left(\frac{1}{B}-1\right)^{\frac{1}{d}-2}
           \left[ 
            \frac{{\rm d}^2B}{{\rm d}F^2} d B(1-B)(d-1-dB)
                    - \left(\frac{{\rm d}B}{{\rm d}F}\right)^2 (d-1)
           \right]\ \ .
\label{square}
\end{align}
\end{widetext}
As the prefactor in the last line is positive we only need the sign
of the term in square brackets $[\ldots]$ 
in order to decide about the sign of the
second derivative. 
%
%
%
Now we use the explicit expression for $B(F)$ in Eq.~\eqref{B(F)}
to calculate the derivatives with respect to $F$.
After some algebra we find for the square bracket in Eq.~\eqref{square}
\begin{align}
   [\ldots]\ =\ & \frac{B(1-B)}{F(1-F)}\ \ \times
\\
                &     \times   
                  \left[
                      -\frac{\sqrt{d-1}}{2\sqrt{F(1-F)}}(d-1-dB)-(d-1)
                  \right]\ \ .
\nonumber
\end{align}
Again, the first factor is positive. Now we substitute the 
expression for $B(F)$, 
Eq.~\eqref{B(F)}, in $(d-1-dB)$ which yields
\begin{align}
  [\ldots]\ =\   \left[-\sqrt{d-1}(d-2)\sqrt{\frac{1-F}{F}}\right]
                  \frac{B(1-B)}{F(1-F)} \ \leqq\ 0\ \ ,
\end{align}
and thus concludes the proof.

\section{Proof of Eq.~\eqref{ineq_distance}}

Consider a family of states $\mathcal{M}$ that are invariant under a group $G$ of entanglement-preserving transformations, that is, $\rho_g := g\rho g^{-1} = \rho$ for $g\in G$ and $\rho\in\mathcal{M}$. Let $\mathcal S^G_k \subset \mathcal{M}$ be the set of states that are symmetric under $G$ and have Schmidt number $k$ ($1\le k<d$), and $\mathcal S_k$ the set of all Schmidt number $k$ states. Given an arbitrary state $\rho$, the minimum distance with respect to the closest Schmidt number $k$ state, $\sigma^\star$, is
\begin{align}
\min_{\sigma\in\mathcal S_k} ||\rho-\sigma||_p &= ||\rho-\sigma^\star||_p = \int dg ||\rho_g-\sigma^\star_g||_p \nonumber\\
&\ge \left|\left|\int dg (\rho_g-\sigma^\star_g)\right|\right|_p = ||\rho_G-\sigma_G^\star||_p \nonumber\\
&\ge \min_{\sigma\in\mathcal S^G_k} ||\rho_G-\sigma||_p \,,
\end{align}
where we have used the triangle inequality in the second line, $X_G:=\int dg g X g^{-1}$, and $||\cdot||_p$ is any Schatten $p$-norm with $1\le p\le \infty$.

%



\begin{thebibliography}{99}
\bibitem{footnote}
We mention that Eq.~\eqref{ineq_distance} holds true for any symmetric family and any appropriate distance measure. See the Appendix for a proof.
\end{thebibliography}
\end{document}